\documentclass[sigconf]{acmart}
\usepackage{amsmath,amsfonts}
\usepackage{algorithmic}
\usepackage{graphicx}
\usepackage{textcomp}
\usepackage{xcolor}
\usepackage{times}  
\usepackage{helvet} 
\usepackage{courier}  
\usepackage{multirow}
\usepackage{graphicx}  
\newcommand{\etal}{{\em et al.}}

\usepackage{makecell}

\newcommand{\BfPara}[1]{\vspace{0.3em}{\noindent\bf#1.}\xspace}
\usepackage{tikz}
\usepackage{booktabs}
\def\BibTeX{{\rm B\kern-.05em{\sc i\kern-.025em b}\kern-.08em
    T\kern-.1667em\lower.7ex\hbox{E}\kern-.125emX}}

\usepackage{blindtext}
\usepackage{booktabs} 
\usepackage{amsfonts}
\usepackage{amsmath} 
\usepackage{booktabs}
\usepackage{multirow}
\usepackage{graphics}
\usepackage{subfigure}
\usepackage{xcolor}
\usepackage{hyperref}
\usepackage{tikz}
\usepackage{enumitem}
\usepackage{booktabs} 
\usepackage{xspace}
\usepackage{makecell}

\usepackage{xcolor,colortbl}

\definecolor{darkgreen}{rgb}{0.0, 0.2, 0.13}
\definecolor{darkred}{rgb}{0.2, 0.0, 0.13}

\newcommand{\acc}[1]{\cellcolor{darkgreen!40!white!#1}}

\definecolor{linkcolour}{rgb}{0,0.2,0.6}
\definecolor{xgreen}{rgb}{0.2,0.6,0.0}
\definecolor{xred}{rgb}{0.7,0.1,0.0}

\AtBeginDocument{%
  \providecommand\BibTeX{{%
    \normalfont B\kern-0.5em{\scshape i\kern-0.25em b}\kern-0.8em\TeX}}}

\copyrightyear{2021}
\acmYear{2021} 
\setcopyright{iw3c2w3}
\acmConference[WWW '21 Companion]{Companion Proceedings of the Web Conference 2021}{April 19--23, 2021}{Ljubljana, Slovenia}
\acmBooktitle{Companion Proceedings of the Web Conference 2021 (WWW '21 Companion), April 19--23, 2021, Ljubljana, Slovenia} 
\acmPrice{}
\acmDOI{10.1145/3442442.3452314}
\acmISBN{978-1-4503-8313-4/21/04}
\author{Sultan Alshamrani}
\email{salshamrani@knights.ucf.edu}
\affiliation{%
  \institution{University of Central Florida}
  \city{Orlando}
  \state{Florida}
  \country{USA}
}

\author{Ahmed Abusnaina}
\email{ahmed.abusnaina@knights.ucf.edu}
\affiliation{%
  \institution{University of Central Florida}
  \city{Orlando}
  \state{Florida}
  \country{USA}
}

\author{Mohammed Abuhamad}
\email{mabuhamad@luc.edu}
\affiliation{%
  \institution{Loyola University Chicago}
  \city{Chicago}
  \state{Illinois}
  \country{USA}
}

\author{Daehun Nyang}
\email{nyang@ewha.ac.kr}
\affiliation{%
  \institution{Ewha Womans University}
  \city{Seoul}
  \country{South Korea}
}

\author{David Mohaisen}
\email{mohaisen@ucf.edu}
\affiliation{%
  \institution{University of Central Florida}
  \city{Orlando}
  \state{Florida}
  \country{USA}
}

\renewcommand{\shortauthors}{Alshamrani et al.}

\begin{document}

\title[Measuring the Exposure of Children to Inappropriate Comments in YouTube]{Hate, Obscenity, and Insults: Measuring the Exposure of Children to Inappropriate Comments in YouTube}

\begin{abstract}
Social media has become an essential part of the daily routines of children and adolescents. Moreover, enormous efforts have been made to ensure the psychological and emotional well-being of young users as well as their safety when interacting with various social media platforms. In this paper, we investigate the exposure of those users to inappropriate comments posted on YouTube videos targeting this demographic. We collected a large-scale dataset of approximately four million records, and studied the presence of five age-inappropriate categories and the amount of exposure to each category. 
Using natural language processing and machine learning techniques, we constructed ensemble classifiers that achieved high accuracy in detecting inappropriate comments. 
Our results show a large percentage of worrisome comments with inappropriate content: we found 11\% of the comments on children's videos to be toxic, highlighting the importance of monitoring comments, particularly on children platforms.
\end{abstract}

\ccsdesc[500]{Computing methodologies~Natural language processing}

\keywords{YouTube Comments, Online Behavior Analysis, NLP.}


\maketitle

\section{Introduction}
The influence of social media on the intellectual and emotional well-being of children and adolescents has been the focus of many studies in recent years, with social media being a central daily activity in children and adolescents' lives alike~\cite{GasserCML12}.
Among the various platforms, YouTube is the most popular video-sharing platform and commonly used by children as an alternative to traditional TV, and as a source of entertainment and educational materials alike. 
A recent study by~\cite{PewInternet2} reported that 81\% of U.S. parents allow their children to use YouTube as an entertainment activity. Moreover, another study shows that children under the age of eight spend 65\% of their time on the Internet using YouTube~\cite{familyzone}.
Therefore, researchers have spent enormous efforts understanding the age-appropriate experience of children and adolescents when using YouTube, and have shown that inappropriate contents---such as contents with sexual hints, abusive language, graphic nudity, child abuse, horror sounds, and scary scenes---are common, with promoters for such contents targeting this demographic~\cite{kaushalSBKP16,abs-1901-07046, TahirASAZW19}. 

Parents and custodians trust children-oriented YouTube channels, such as Nick Jr., Disney Jr., and PBS Kids, to present educational and entertaining material for their children even with no supervision. 
However, children can be exposed to inappropriate and disturbing videos, suggested by the YouTube recommendation system, as children are tricked to click on innocent-looking thumbnail~\cite{abs-1901-07046}.
To ensure their well-being and safety, it is important to study the exposure of children and adolescents to inappropriate material presented on YouTube, including visual, audio, and written content. Even when watching videos from trusted family-friendly channels, the written contents, such as user comments, might contain inappropriate language that could influence the children's offline behavior. The limited work on YouTube textual contents, as opposed to the various efforts on understanding YouTube's video/audio contents, creates the need for comments-based studies.

Our study explores measuring the exposure of children and adolescents to age-inappropriate comments posted on videos of the top-200 children shows~\cite{Ranker}. This task is challenging for several reasons. First, studying comments on children videos requires manually collecting channels and shows targeting this demographic, knowing YouTube categories are not established by age-group but rather by the topic they present. Second, assigning age groups to the collected videos can be daunting in measuring exposure by separate groups. Third, the lack of a ground truth dataset for safe and inappropriate content posted on such videos makes it difficult for machine-learning models to capture the children's exposure on a large scale.
Considering the variety of age-inappropriate content for children, building a unified system for detecting such contents is challenging.

To address those challenges, we built a large collection of YouTube comments on children-oriented videos for the top 200 shows categorized by different age groups~\cite{CommonSenseMedia}. We extended the dataset with ground truth data from different sources to establish five age -inappropriate categories; toxic, obscene, insult, and identity hate.
The used ground truth dataset compresses annotated data provided by Conversation AI on Wikipedia's comments, and our own manually-annotated data from YouTube comments posted on children videos. We leveraged natural language processing and machine learning techniques to construct an ensemble of models, each of which specializes in detecting a specific inappropriate category. The models are trained and tested on ground truth samples, and separately and collectively achieve remarkable results. Utilizing our ensemble, we uncovered a large number of age-inappropriate comments among those posted on children YouTube videos. Measuring the exposure by age group, our results show that children between 13 and 17 years old are the most exposed to such contents. For inappropriate categories, toxic-related comments are the most common, with 15.54\% out of the total comments, then insult (7.96\%) and obscene (6.84\%). 

\BfPara{Contribution}
This work contributes to measuring the exposure of children to inappropriate content present in the kids' YouTube videos comments. We summarize our contribution as follows: 
\begin{itemize}
    \item We collected a large-scale dataset of comments on children's YouTube videos from the top-200 ranked children shows. The list of shows, retrieved search results, categorization of shows by age group, and other artifacts related to the data collection process are manually vetted. 
    \item We built a manually-annotated ground truth dataset collected from comments posted on children's videos, which includes about 6,000 comments. 
    \item Leveraging natural language processing and deep learning techniques, we designed and implemented an ensemble of classifiers to detect five age-inappropriate contents. Models of the ensemble are trained, fine-tuned, and evaluated using the ground truth dataset.
    \item Adopting the ensemble classifier on the YouTube comments domain, we detected and measured children's exposure to inappropriate comments.
    \item We provided an in-depth analysis of children's exposure to inappropriate content in terms of age groups, user interactions, and YouTube video channels.
\end{itemize}

\section{Related Works}\label{sec:related_works}

Recently, several studies have been conducted with the aim of exploring the effects of social media on children, since the use of social media has become a significant part of their daily routines. To ensure the safety of kids on YouTube, Alshamrani~\etal~\cite{AlshamraniAM20} studied the exposure of children to malicious URLs on videos targeting young users. Another study by~\cite{OGCK2011} which has encouraged parents to understand and be aware of the various possible offline and online behaviors of their children, such as cyber-bullying, privacy issues, sexting, and Internet addiction.
Among other social media platforms, YouTube has been the subject of many studies since it is considered the most popular social media platform in the United States~\cite{PewInternet}, and the second-largest search engine after Google worldwide~\cite{MushroomNetworks}.
Studying the appropriateness of contents being presented to children on YouTube was first considered, to the best of our knowledge by Kaushal~\etal~\cite{kaushalSBKP16} who studied kids-unsafe contents and promoters. The authors provided a framework for detecting unsafe contents using measures calculated on the video, user, and comment levels with an accuracy of 85.7\%.

Another work by Papadamou~\etal~\cite{abs-1901-07046} shows that inappropriate toddler-oriented videos are common and likely to be suggested by YouTube's recommendation system. 
Using manually-annotated videos, the authors investigated the detection of inappropriate content (containing sexual hints, abusive language, graphic nudity, child abuse, horror sounds, and scary scenes) collected from videos targeting kids using deep learning algorithms to achieve an accuracy of 84.3\% for this task.  
More recently, Tahir~\etal~\cite{TahirASAZW19} demonstrated that even children-focused apps, such as {\em YouTube Kids} which is considered a kids-safe platform, are prone to compromise with inappropriate videos. 

As part of studying users' comments, Alexandre~\etal~\cite{CunhaCP19} studied and analyzed users' opinions on several aspects, such as the quality of the video, YouTuber presence, and videos' contents.
Improving the content enables achieving higher popularity as Figueiredo~\etal~\cite{FigueiredoABG14} outlined. In particular, the quality and user perception of the contents facilitate popularity on YouTube. Bermingham~\etal~\cite{BerminghamCMOS09} is another related work, in which they provided YouTube comment-based sentiment analysis of topics potentially serving a radicalizing agenda. Recently, several work explored the analysis and detection of hate speech in different social media platforms~\cite{ZannettouEBNS20, thomas2021sok,MaricontiSBCKLS19, AlshamraniAAM20}. This work studies and measures the exposure of children to age-inappropriate content in YouTube comments posted on children's shows.

\begin{figure*}
\centering
\begin{minipage}[t]{0.32\textwidth}
\includegraphics[width=0.99\textwidth]{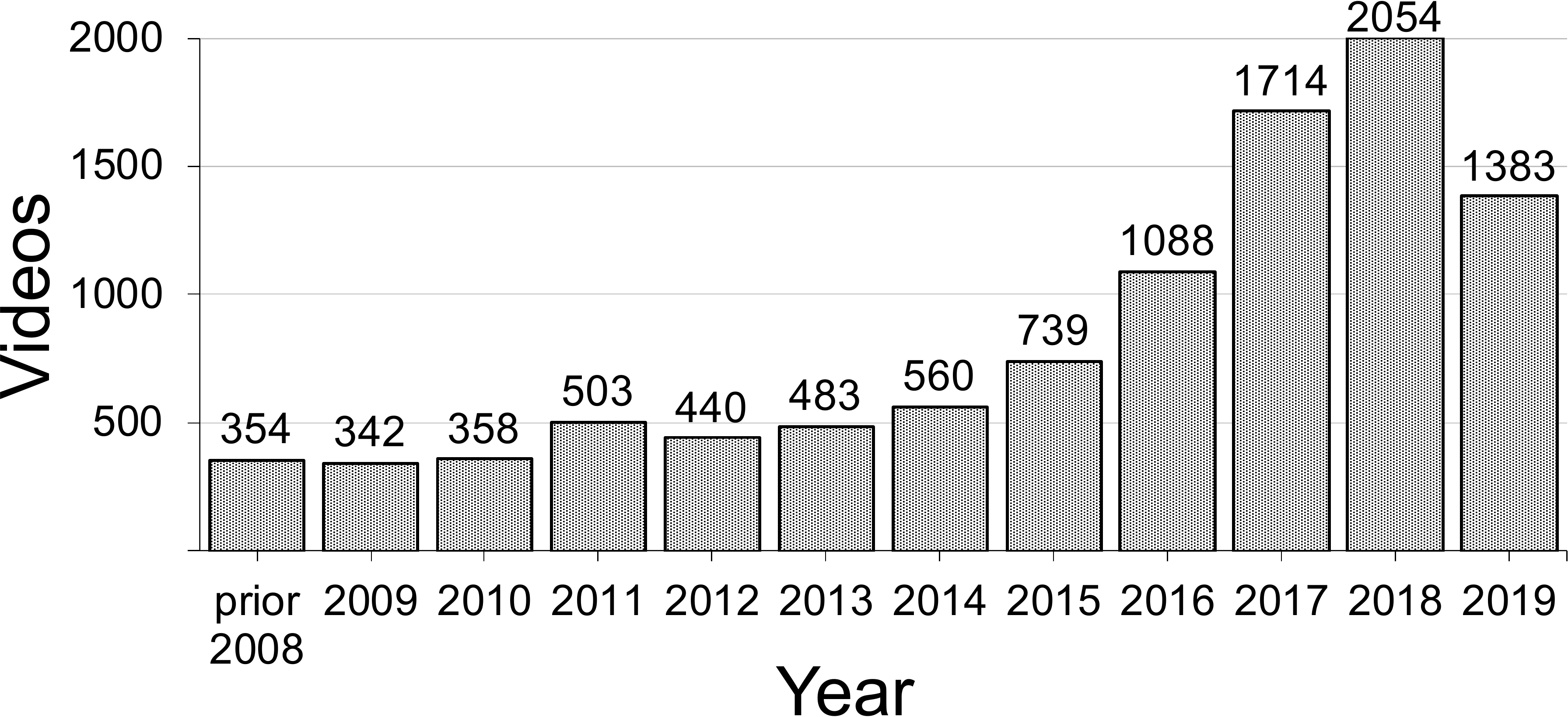}\vspace{-2mm} 
\caption{The publish date distribution of the collected YouTube kids' videos.}
\label{fig:videos_years}
\end{minipage}
\hfill
\begin{minipage}[t]{0.32\textwidth}
\includegraphics[width=0.99\textwidth]{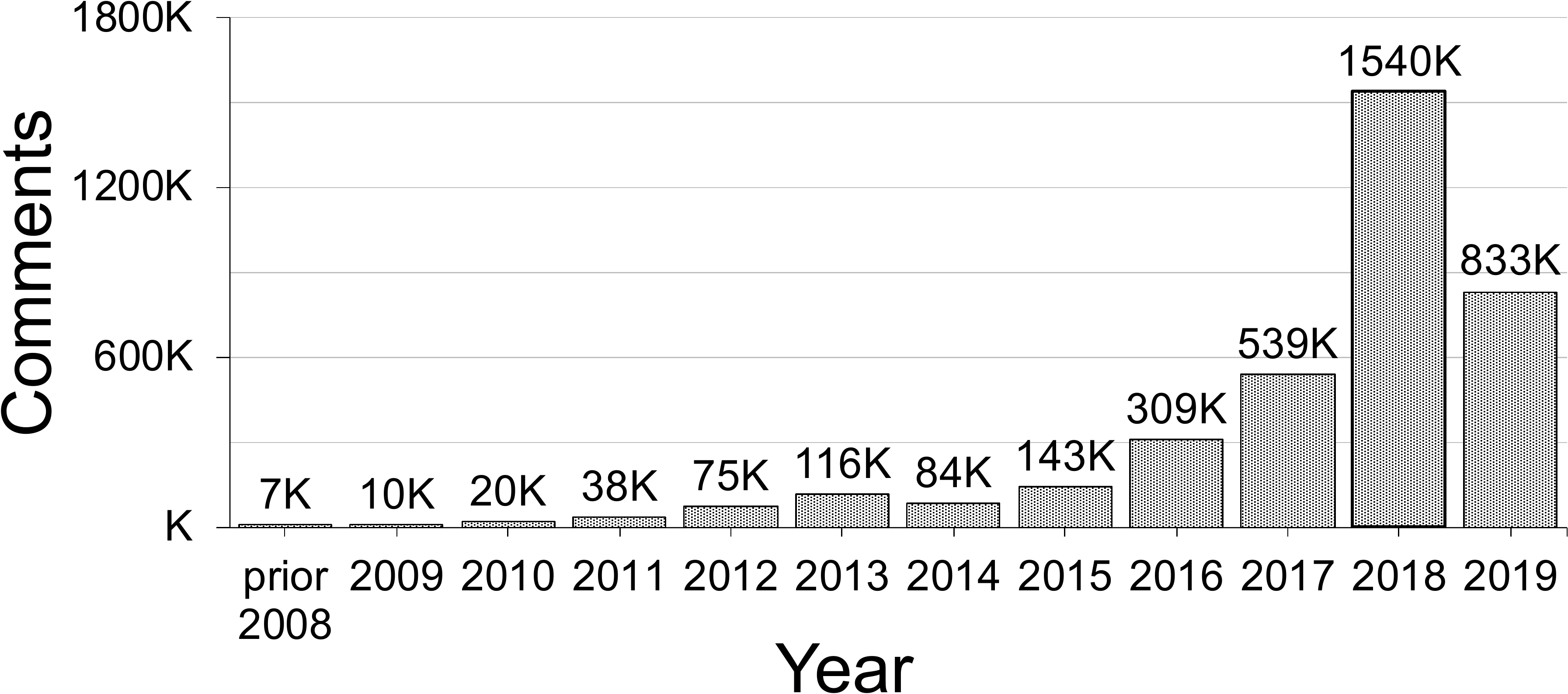}\vspace{-2mm} 
\caption{The distribution of YouTube kids' videos comments over past years.}
\label{fig:comments_years}
\end{minipage}
\hfill
\begin{minipage}[t]{0.32\textwidth}
\includegraphics[width=0.99\textwidth]{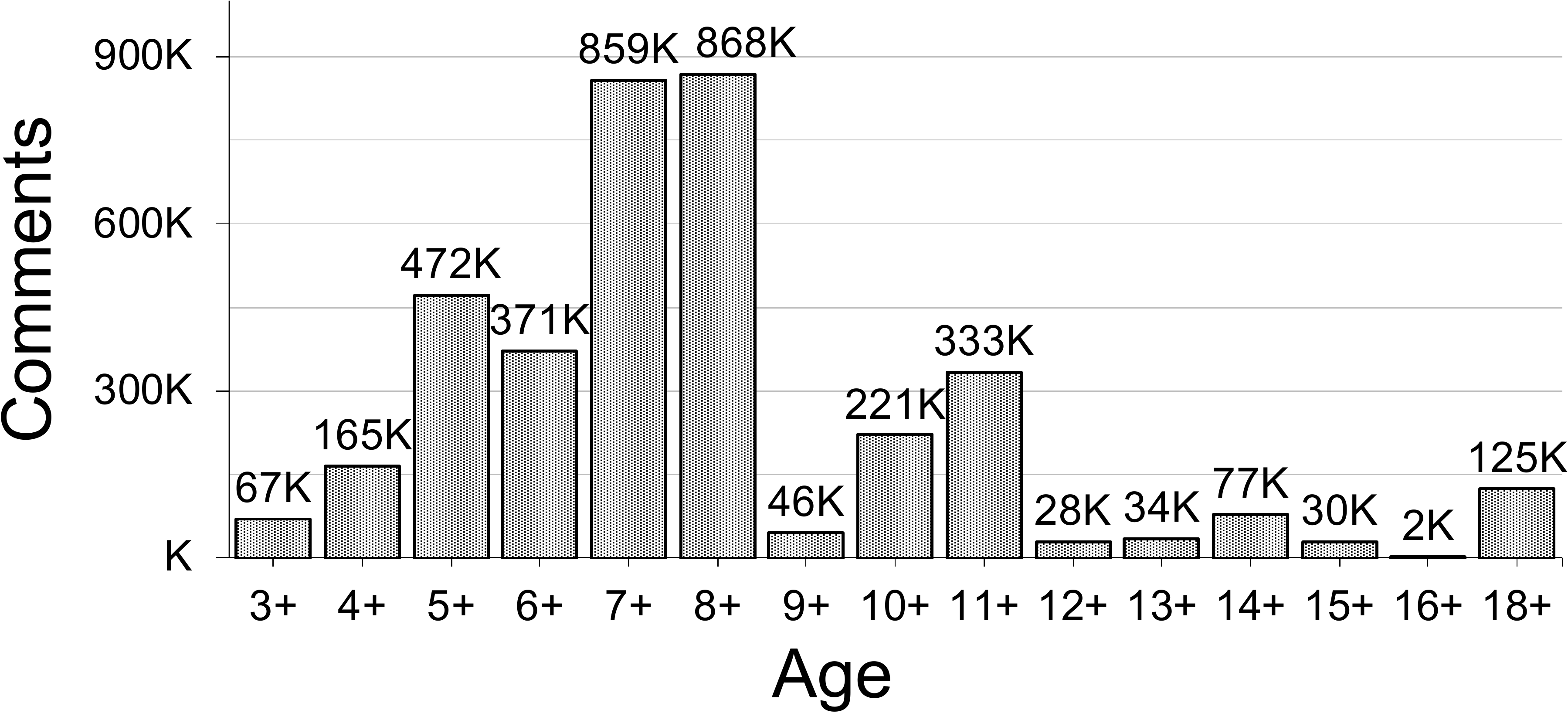}\vspace{-2mm} 
\caption{The distribution of YouTube kids' comments over different ages.}
\label{fig:age_comments}
\end{minipage}%
\vspace{-2mm} 
\end{figure*}

\section{Approach and Techniques}\label{sec:dataset}

\begin{table}
\centering
\caption{The distribution of the collected dataset. The collected comments are from two sources: Wikipedia and YouTube. 5,940 YouTube comments are manually labeled for the evaluation of the ensemble models.}\label{tab:dataset}\vspace{-2mm}
\begin{tabular}{l|l|r}
\Xhline{2\arrayrulewidth}
{\bf Source}                     & {\bf Dataset}       & {\bf Count}       \\
\Xhline{2\arrayrulewidth}
\multirow{6}{*}{\bf Wikipedia~~~~~~~~} & Safe Comments~~~~~~~~ & ~~~~~~~~143,000     \\
                          & Toxic         & 15,294      \\
                          & Obscene       & 8,449       \\
                          & Insult        & 7,877       \\
                          & Threat        & 478         \\
                          & Identity hate & 1,405       \\
                          \Xhline{2\arrayrulewidth}
\multirow{7}{*}{\bf YouTube}   & Unlabeled     & $\approx$~3,700,000 \\
                          & Safe Comments & 1,832       \\
                          & Toxic         & 4,126       \\
                          & Obscene       & 2,367       \\
                          & Insult        & 1,650       \\
                          & Threat        & 550         \\
                          & Identity hate & 788        \\
                          \Xhline{2\arrayrulewidth}
\end{tabular}\vspace{-4mm}
\end{table}

\subsection{Data Collection and Measurements}
Our dataset includes YouTube comments and two datasets of ground truth, one from the Conversation AI team and another one annotated by our team for ground truth from the YouTube comments. For the YouTube comments, we collected more than 3.7 million comments posted on roughly 10,000 children's videos, distributed over the period from January 2005 until March 2019.

\BfPara{Children's Shows}
We collected comments on videos of the top-200 children's shows based on Ranker~\cite{Ranker}, a crowdsourced platform that relies on millions of users to rank a variety of media contents such as shows and films. 
The list of shows was originally made by Ranker TV and received more than 1.2M votes, and has 380 kids' shows. Among them, we selected the top 200 shows. We augmented our list with part of Wikipedia's list of cartoon shows.

\BfPara{Collection Approach} 
Using YouTube APIs, we extracted the top-50 videos of the search results on every show on our list. Using each retrieved video's ID, we also used the API to obtain video statistics, such as the number of views, likes, dislikes, etc.
We used YouTube Comments API to collect all comments from the videos. 
In total, we collected more than 3.7 million comments from 10,000 videos. 

\BfPara{Age-Appropriateness of Children's Shows} 
We defined age appropriateness as the adequate age group to be the subject of the show. 
Defining the age appropriateness for children's shows is challenging, since most shows do not specify the target age group. Therefore, we used  {\em Common Sense Media}~\cite{CommonSenseMedia}, a non-profit organization that provides education and advocacy to families on providing safe media for children, as the main source for defining the age group of the targeted children's shows.  
Using {\em Common Sense Media}, we were able to retrieve the appropriate age group for most of the kids' shows on our list. However, a few shows do not appear in {\em Common Sense Media}, and for those we turned to IMDB~\cite{imdb}, an online database of information about different types of media such as films, television programs, home videos, video games, etc., to obtain the age group for those particular shows.  
Some shows have different versions, each for a certain age group, therefore the age group is assigned based on the most prevalent version in the YouTube search. 
Some other kids shows are assigned an age group based on their respective categories, e.g., Loony Tunes (a well-known collection of cartoons for age 7+).    
We note that we conducted a manual inspection on the age appropriateness for the retrieved top-50 results on each show to define non-kids contents and assigned them to 17+ age group, which is the highest age group in our dataset.

\BfPara{Data Statistics and Measurements} Here we provide general statistics of our data. 
The collected YouTube comments were posted by more than 2.5 million users on about 10,000 videos from more than 3,000 different channels.   
These retrieved videos have an average viewers count of roughly 2.4 million views and an average comments count of 8,068 comments per video. 
Observing the publishing date of the videos in our collection, Figure~\ref{fig:videos_years} demonstrates the rapid increase in children's videos over the past few years. The figure shows an increase in popularity of five folds in ten years from 2008 (with 354 videos) to 2018 (with 2,054 videos).  
This rapid growth in popularity is observed through the first three months of 2019 with 1,383 videos included in our collection (by March of 2019).
We note that the collection of YouTube videos is based on their relevance, and not the publishing date nor the view count; this is also the case when retrieving videos from the top-50 search result and when querying the targeted shows.
The search results do not always reflect the popularity. However, the top-ranked videos are often characterized by bursts of popularity~\cite{FigueiredoBA11}.
Generally, a consistent trend is observed in the year-over-year increasing number of videos included in our collection.
Similar patterns are observed with the number of comments from around 7,000 comments on videos prior to 2008 to more than 1.5 million comments on videos from 2018. This growth is steady through the first three months of 2019 as illustrated in~\autoref{fig:comments_years}.
We also provided the distribution of comments across the age groups as shown in~\autoref{fig:age_comments} where most of the collected comments were posted on videos for kids between the age of five and eight (a total of approximately 2.5 million comments). 
 
\BfPara{Age-Appropriateness of Contents} 
Contents that are regarded as age-appropriate for children and adolescents ideally should not contain toxic words or imply an insult, threat, identity hate, or obscenity. To study the appropriateness of YouTube comments, we collected ground truth datasets to establish a baseline for modeling contents with different labels (i.e., toxic, obscene, insult, threat, and identity hate). The ground truth data includes: (1) labeled comments from Wikipedia that is manually annotated by Conversation AI, a research team started by Jigsaw and Google to provide tools and solutions for improving online conversions; (2) labeled comments posted on YouTube videos targeting children that are manually annotated for the purpose of this study.

\BfPara{(1) Wikipedia Ground Truth Toxic Dataset}
We used the manually-annotated dataset provided by Conversation AI, with approximately 160,000 comments from Wikipedia Talk pages of which approximately 143,000 comments are labeled as safe, while the remaining are labeled to have different types of toxicity (i.e., 15,294 toxic, 8,449 obscene, 478 threat, 7,877 insult, and 1,405 identity hate). A summary of the collected data is provided in~\autoref{tab:dataset}.  

\BfPara{(2) Manually Annotated Ground Truth} We manually annotated 5,958 YouTube comments posted on YouTube videos for the evaluation of the ensemble. The total number of the manually labeled comments is distributed as follows: safe: 1,832, toxic: 4,126, obscene: 2,367, insult: 1,650, threat: 550, and identity hate: 788. 

For our manual labeling, we used several explicit rules. Each comment was labeled as either toxic or safe. A toxic comment may belong to one or more of unsafe category; obscene, threat, insult, or identity hate. A comment is considered obscene when it is morally offensive in a sexual way, or when it has socially offensive words. When such an offensive language is used against or to describe other users, video publishers, or anyone else, the comment is considered as an insult. When such an offensive language is directed to another group of people, by imposing a negative stereotype or prejudices about people based on their race, color, or ethnicity, the comment is considered as an identity hate. 

The annotation is challenging since identifying identity hate is highly subjective~\cite{OlteanuTV17}~\cite{RossMGBNM17}. Some comments did not have any profanity or offensive language, but implied a threat to other users or the video publisher; we labeled such a comment to be a threat. In the annotation process, we encountered comments that are socially unacceptable and are age-inappropriate, however, they do not belong to any of the four unsafe categories, and so we labeled them as {\em toxic} only. The manual labeling has been done by the same annotator (lead author of this work), upon refining the above ruleset. We avoided using multiple annotators across different folds of the manually-labeled dataset, and rather pursued this slow labeling method, to avoid inconsistency and subjectivity in interpretation against the predetermined labeling rules. 

\BfPara{Ground Truth: Safe Dataset}
For safe content, we used our labeled safe YouTube comments as well as safe-labeled comments from the Conversation AI team dataset, which include roughly 143,000 comments in total.

\begin{figure*}[t]
\centering
\includegraphics[width=0.95\textwidth]{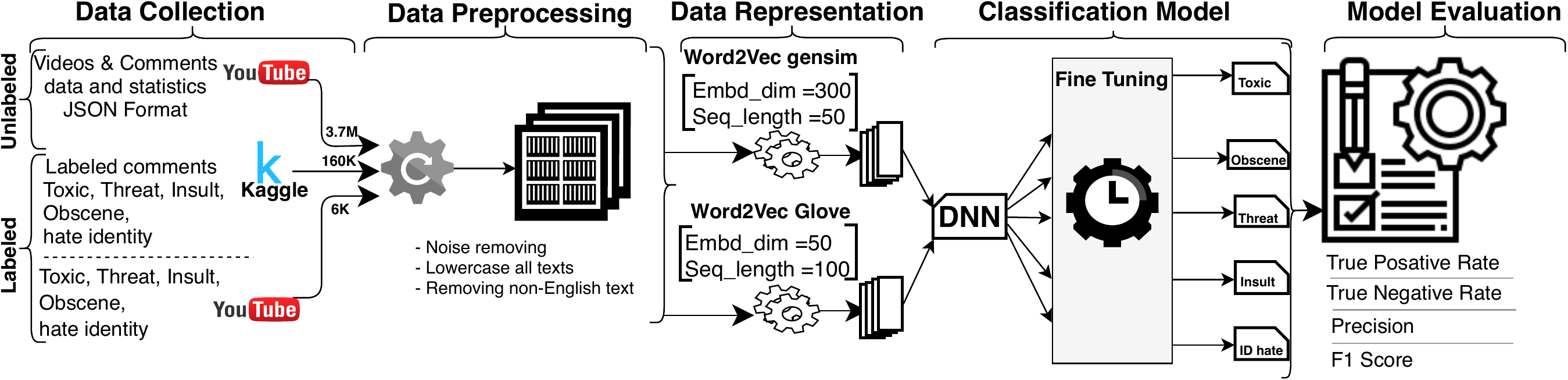}\vspace{-3mm} 
\caption{The ensemble pipeline. The system design consists of five stages, starting from data collection and labeling, followed by the preprocessing of the data to generate efficient representation. Then, ensemble of five classification models are used for comments classification. Further, the models are evaluated using four evaluation metrics.}\label{fig:system_pipline}\vspace{-2mm} 
\end{figure*}

\subsection{Data Preprocessing}\label{sec:preprocessing}
Several preprocessing steps are taken before the final data representation, modeling, and evaluation, and to ensure clean and proper representation of the collected data. YouTube comments are  the focus of this study, which we  addressed with the preprocessing steps as follows: (1)
We initially removed all {\em non-English contents} across all datasets, and limit our analysis to English comments. 
(2) We eliminated unwanted characters and tokens, e.g., punctuation, and other characters that represent or encode emojis.

\subsection{Data Representation}
\BfPara{Comments Data Representation}
In order to perform an analysis of textual data, we first transformed this data into an embedding (i.e., numerical representation) that can be used by machine learning models. Such a representation allows the machine learning models to learn and capture different patterns of the text. We utilized different data representation methods, namely, \textit{Word2Vec}~\cite{MikolovCCD13} and \textit{Glove}~\cite{PenningtonSM14}.

\BfPara{Pre-trained Word2Vec} 
Using the pre-trained model for comments representation, 
we have the following two cases of distinct models. \textbf{(1) Gensim:} This technique transforms textual data by examining word statistical co-occurrence patterns within a corpus of the provided textual documents. Examining different configurations for both word embedding and the document vector. We found that the highest accuracy can be achieved using a size of 300 for the word embedding and the document vector size of 50. \textbf{(2) Glove:} This technique is an unsupervised learning algorithm used to generate numerical vector representations for words. The training process is done on aggregated global word-word co-occurrence statistics from a corpus. We used Glove to represent the comments; similar to Gensim, we tried different configurations and selected the configuration with the highest accuracy, using a size of 50 for the word embedding and 100 for the document vector.

\subsection{Ensemble Classification Models}
To understand and measure children's exposure to inappropriate comments on YouTube videos by first identifying them, we adopted an ensemble classifier to build five specialized models for classifying five unsafe categories: {\em toxic}, {\em obscene}, {\em threat}, {\em insult}, and {\em identity hate}. The models are trained, in a supervised manner, using the Wikipedia toxic comments dataset and the manually annotated ground truth of YouTube comments. Each model predicts whether an input belongs to a specific category, functioning as a binary classification task. We note that a comment can belong to one or more categories (e.g., toxic, insult, and identity hate simultaneously), thus the output of the ensemble is positive if the comment is labeled as at least one age-inappropriate category. 

Our approach adopts an ensemble of classifiers to predict different age-inappropriate categories using DNN models. Based on our experiment, DNN performs very well in terms of identifying different age-inappropriate categories as opposed to CNN and RNN. We found that different pre-trained models for feature representation, such as Glove and Gensim, work better in certain scenarios for identifying certain age-inappropriate comments categories (i.e., Glove with DNN for identifying threat comments). The ensemble uses DNN for identifying five age-inappropriate categories, DNN with gensim Word2Vec for identifying toxic, obscene, insult, and identity hate categories, and DNN with Glove Word2Vec for identifying threat category.

We feed Word2Vec vectors of the comments to the first input layer in the network while the output layer has a single node for binary classification to predict whether the provided comment belongs to a certain class or not. Our model architecture is composed of two dense layers of size 128 units with a ReLU activation function, each followed by a dropout operation with a rate of 20\%. The last layer is fully connected to a sigmoid function, which generates real values in the range (0,1) using the function ${sigmoid}(z)={1}/{(1 + e^{-z})}$.
Since the output $\{y \in \mathbb{R} ~|~ 0 \leq y \leq 1\}$, determines the probability of assigning an input to the target, a threshold can be defined for target $\bar{y}$ assignment (e.g., a commonly-used threshold is $0.5$ where $\bar{y}=1 ~if~ y \geq 0.5$). We explored different thresholds for each category to optimize the true negative rate and the true positive rate.

\BfPara{Model Training Settings} We used the entire Wikipedia annotated comments to train five models, each of which is specialized in detecting one age-inappropriate category. Then we fine-tuned the trained model using 50\% of our manually labeled comments from YouTube, by only retraining the last layer of the model. We then used the other half for the evaluation of the models. The training process is guided by minimizing the binary-cross-entropy as follows:
\[\textrm {loss}(\theta) =  \frac{-1}{N} \sum_{i=1}^N[y_i \times \log (p_i) + (1 - y_i) \times \log(1 - p_i)],\]
where $p_i$ is the conditional probability $p(y_i | x_i, \theta)$ for a target $y_i$ given an input $x_i$ and a set of parameters $\theta$, $i$ is the $i$-th record, and  $N$ is the total number of records in the training set. The optimization is done using {\em RMSprop} optimizer, a stochastic optimization algorithm, with a learning rate of $10^{-3}$ without decaying over time. We used a mini-batch approach with a batch size of $128$, and for preventing the overfitting we used {dropout} regularization with a dropout rate of $0.2$. The termination criterion is set to be a specified number of training iterations, which is set to 100 for all models.

\BfPara{Evaluation Metrics}
This study uses four evaluation metrics, which are {\em Precision}, {\em F1-score}, {\em True Positive Rate} (TPR), and {\em True Negative Rate} (TNR). 
{Precision} represents the percentage of which a model was correct in predicting the positive class (P = $\text{TP}/\text{TP+FP}$). {F1-score} is the harmonic mean of the precision and recall, and is expressed as ($\text{F1-score} = \text{2TP}/\text{2TP+FP+FN})$ where TP, FP, and FN represent True Positive, False Positive, and False Negative, respectively.
The {TPR} is the proportion of the positive predictions, positive labeled-data correctly predicted to be positive, form the total positive-labeled data ($\text{TPR}= \text{TP}/\text{TP+FN}$). The {TNR} is the proportion of the negative predictions, negative labeled-data correctly predicted as negative, from the total of negative-labeled data ($\text{TNR}=\text{TN}/\text{TN+FP}$).

\begin{figure*}[ht]
    \subfigure[Toxic\label{fig:toxic_threshold_R}]
    {\includegraphics[width=0.196\textwidth]{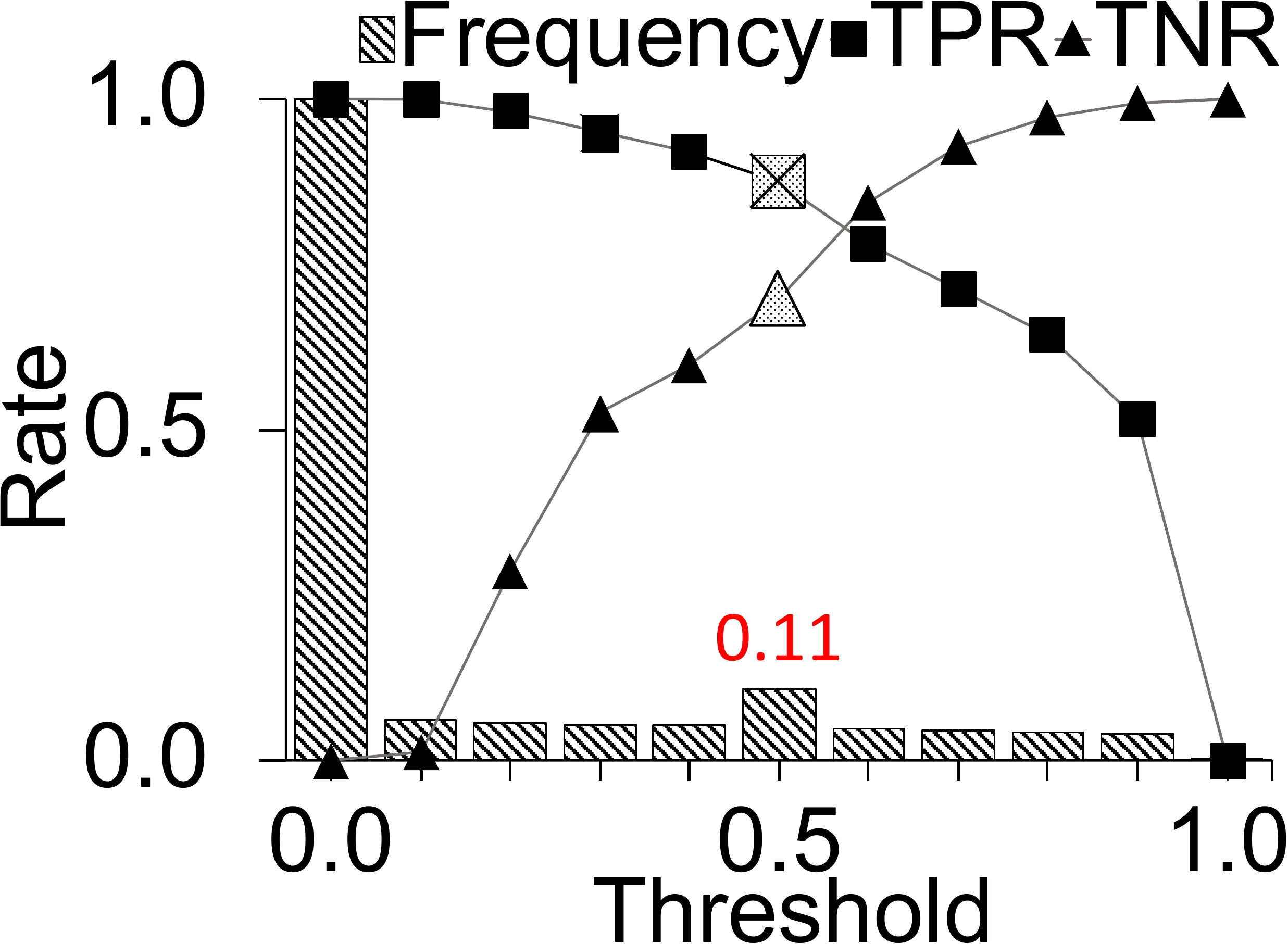}}
    \subfigure[Threat\label{fig:threat_R}] {\includegraphics[width=0.196\textwidth]{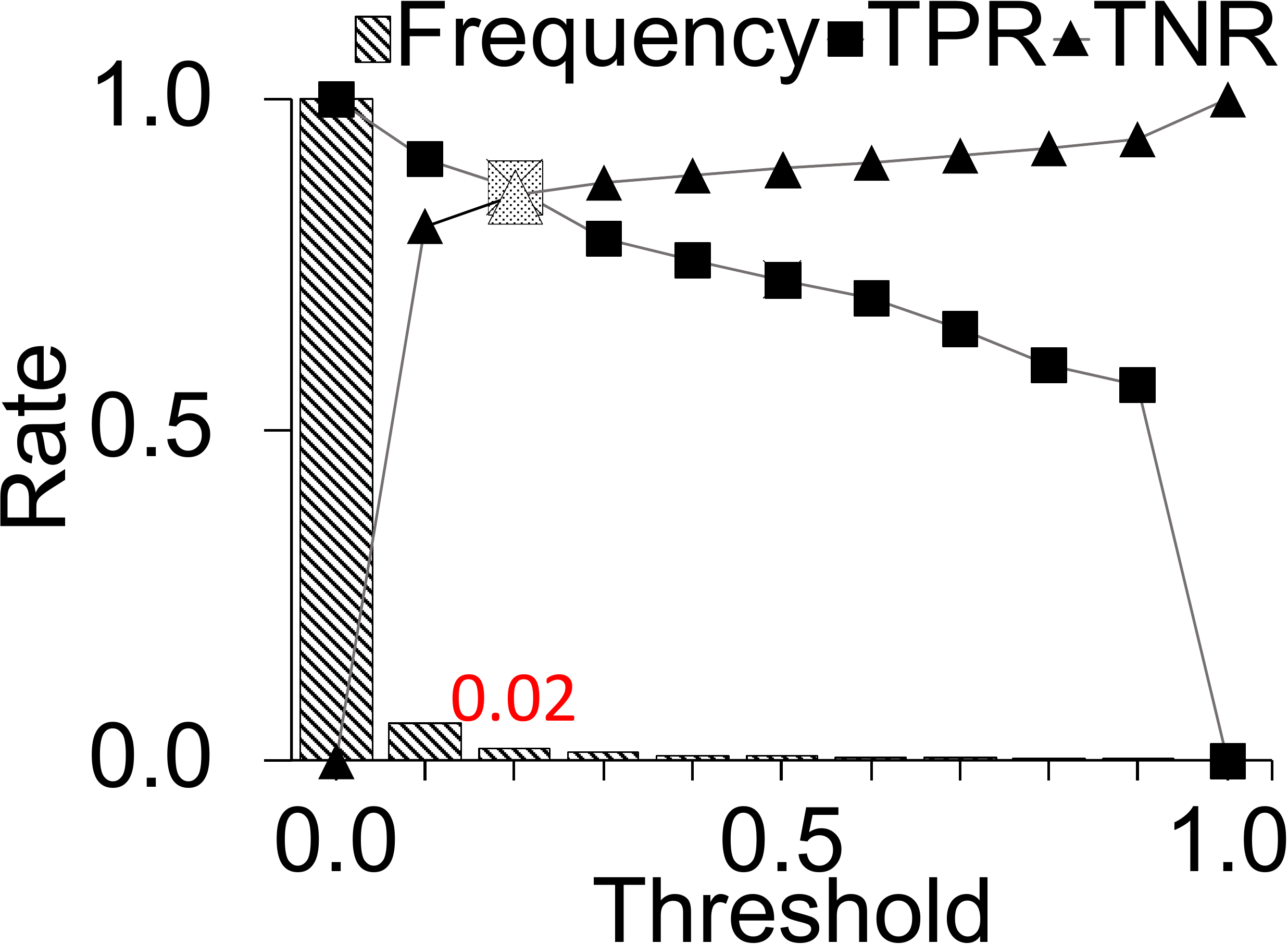}}
    \subfigure[Insult\label{fig:insult_R}] {\includegraphics[width=0.196\textwidth]{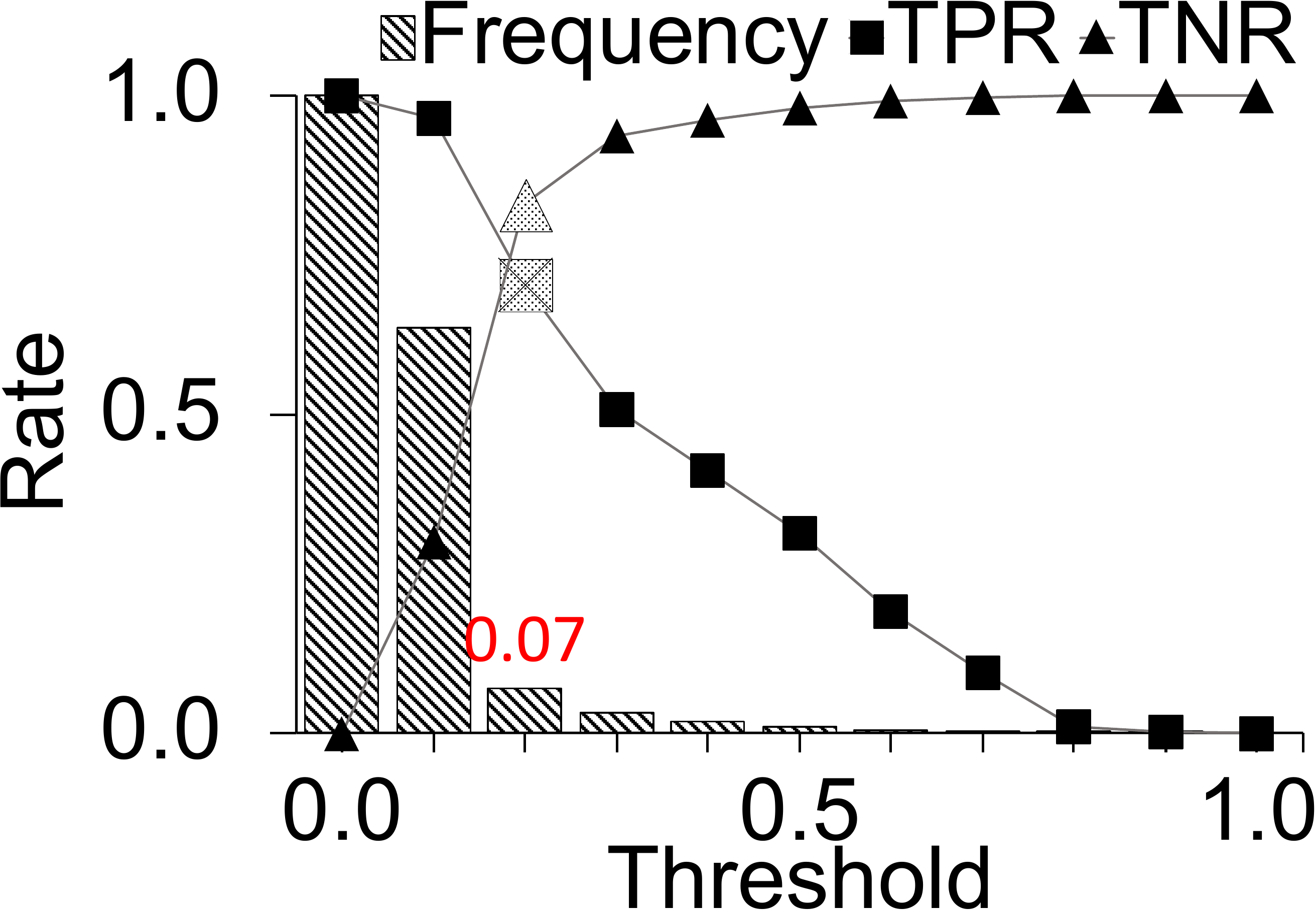}}
    \subfigure[Obscene\label{fig:obscene_threshold_R}] {\includegraphics[width=0.196\textwidth]{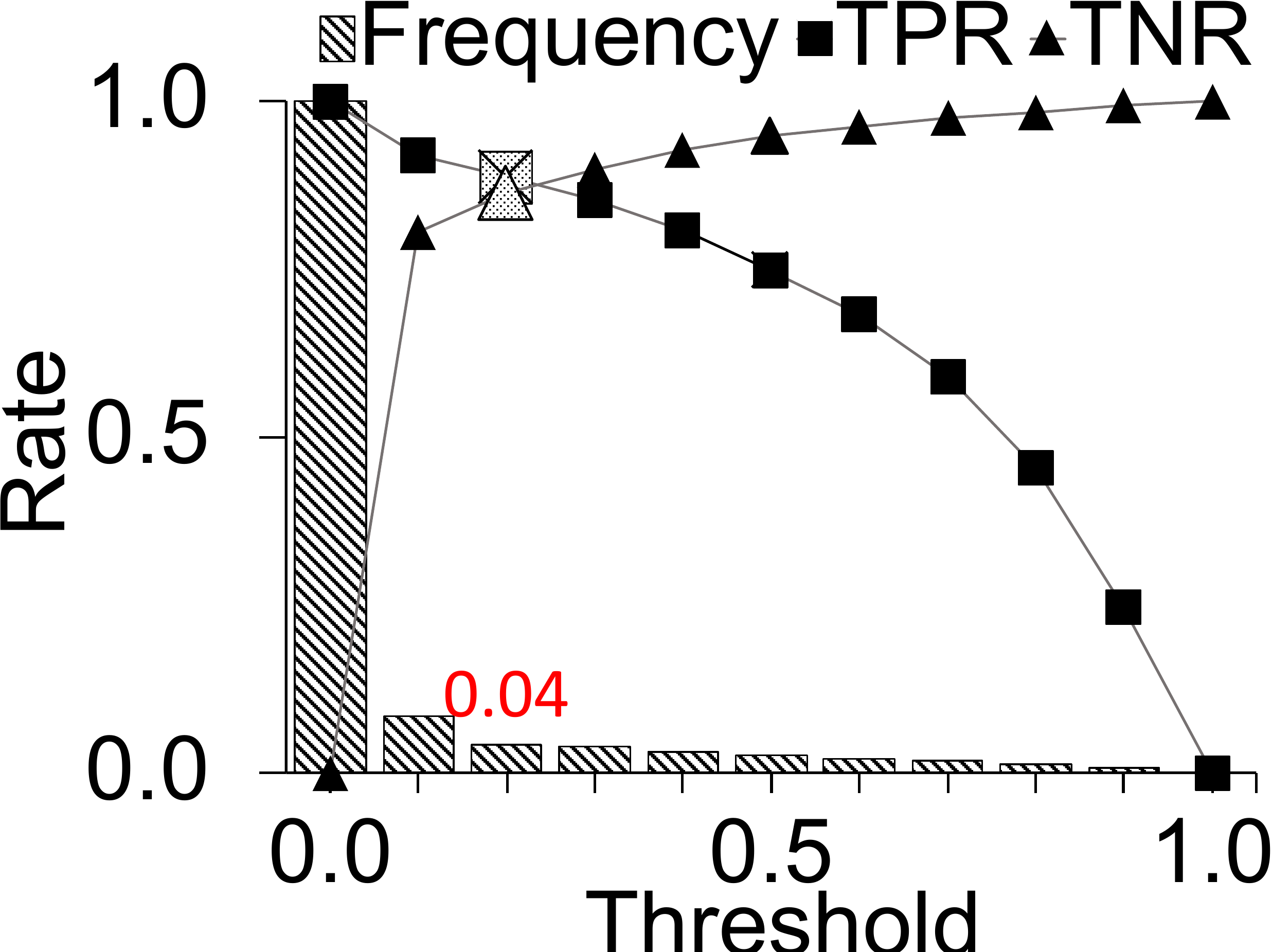}}
    \subfigure[Identity hate\label{fig:identity_hate_R}] {\includegraphics[width=0.196\textwidth]{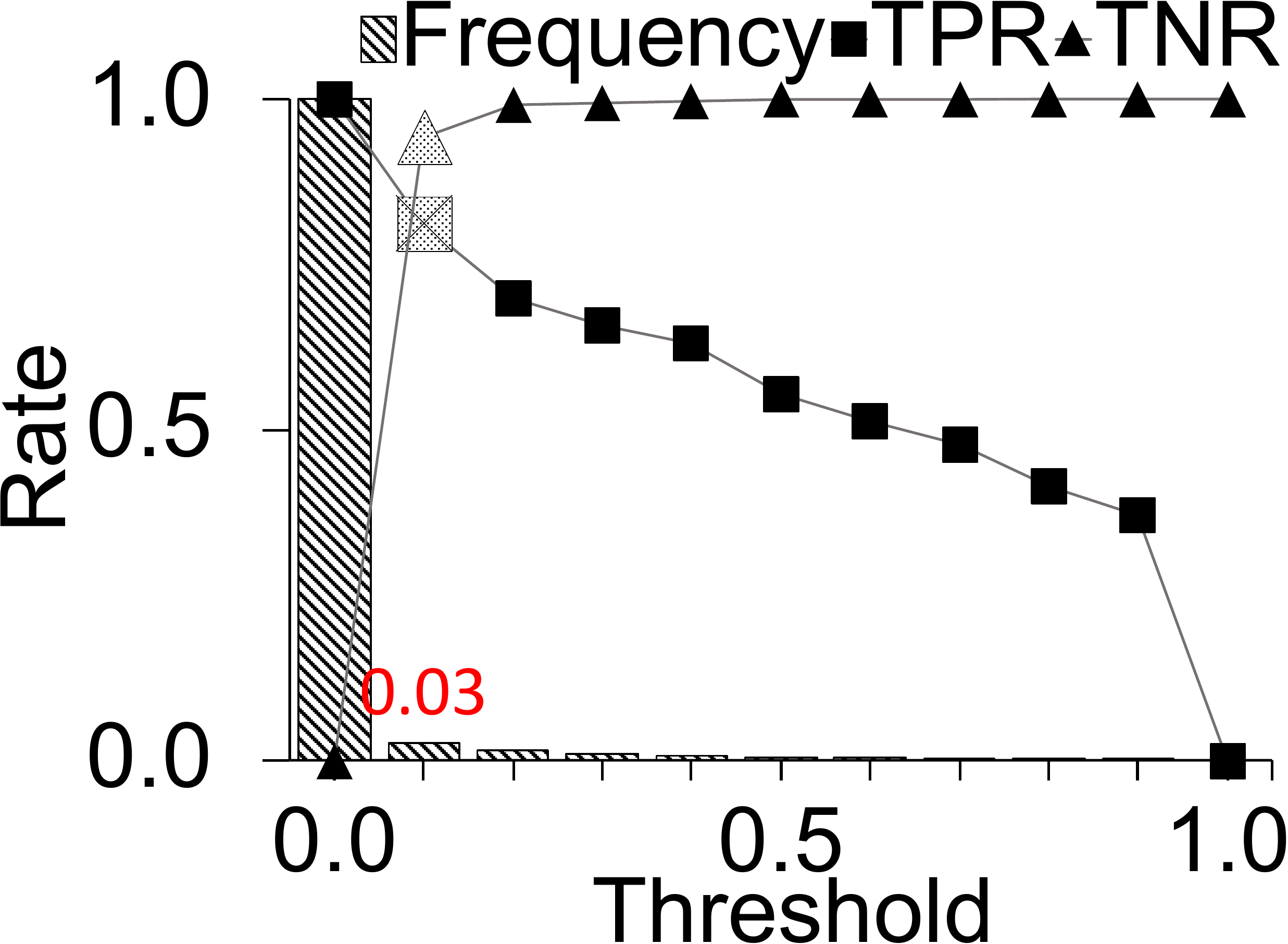}}\vspace{-2mm} 
    \caption{The evaluation of the ensemble model across categories in terms of TPR and TNR. The x-axis represents the chosen threshold, and y-axis shows the respective TPR, TNR, and percentage of detected YouTube comments.}\label{fig:thresholds}\vspace{-3mm} 
\end{figure*}

\begin{table}[t]
\centering{
\caption{The performance of the ensemble model on Wikipedia comments, and the fine-tuned models across different metrics. Overall, the fine-tuned ensemble achieved a TNR of 86.8\% and TPR of 78.5\%.}\label{tab:performance}\vspace{-2mm} 
\begin{tabular}{l|c|c|c|l|l|l}
\Xhline{2\arrayrulewidth}
   \multirow{2}{*}{\bf Class}           & \multicolumn{3}{c|}{\bf Wikipedia} &\multicolumn{3}{c}{\bf Fine Tuned} \\
   \cline{2-7}
         & Recall & Prec & F1            & Recall  & Prec & F1            \\
\Xhline{2\arrayrulewidth}
\bf {Toxic}         & \acc{92}92.5   & \acc{83}82.5 & \acc{88}87.2 & \acc{94}93.5    & \acc{83}83.1 & \acc{88}88.0 \\
\bf {Obscene}       & \acc{82}81.9   & \acc{83}82.9 & \acc{82}82.4 & \acc{87}86.6    & \acc{84}83.5 & \acc{85}85.0 \\
\bf {Threat}        & \acc{64}64.4   & \acc{44}43.7 & \acc{52}52.1 & \acc{71}71.3    & \acc{42}42.3 & \acc{53}53.1 \\
\bf {Insult}        & \acc{75}74.5   & \acc{56}55.7 & \acc{63}63.3 & \acc{67}66.7    & \acc{64}64.4 & \acc{66}65.6 \\
\bf {Identity hate} & \acc{54}53.9   & \acc{90}89.8 & \acc{67}67.4 & \acc{75}74.8    & \acc{88}87.8 & \acc{81}80.8 \\
\Xhline{2\arrayrulewidth}
{\bf{Overall}} & \acc{74}73.4 & \acc{71}70.9 & \acc{70}70.4 & \acc{76}78.5 & \acc{72}72.2 & \acc{75}74.5
\\
\Xhline{2\arrayrulewidth}
\end{tabular} }\vspace{-2mm} 
\end{table}

\section{Results and Discussion}\label{sec:ResultsAndDiscussion} 
In this section, we review the results of the ensemble for classifying five categories of inappropriate contents, including, {toxic}, {obscene}, {threat}, {insult} and {identity hate}. 
Then, we measured children's exposure to inappropriate comments on YouTube using the best-performing models.

\subsection{Ensemble Model Performance}
The ensemble model performance is reported in~\autoref{tab:performance} using three metrics. We reported the performance of the models trained on Wikipedia then evaluated on the annotated YouTube comments as well as the performance of these models after being fine-tuned. The results are based on the specific probability threshold providing the best trade-off between TPR and TNR as shown in~\autoref{fig:thresholds}. An emphasis on high TPR is considered when choosing the threshold to ensure high correctness for positively predicted output (i.e., some positive contents might not be detected but barely mistaken when they are detected). 
This high performance can be seen with the F1-score, with a high of 86.6\% for the toxic and a low of 52.9\% for the threat. 
We also observed the challenge in achieving high TPR for the threat and identity-hate categories due to several reasons, including the limited number of samples for those categories (see~\autoref{tab:dataset}) and the ambiguity caused by the used language.

\subsection{Ensemble Adoption and Measurement}
Using the best TPR-TNR trade-off thresholds, 
we constructed an ensemble model to evaluate and measure kids exposure to inappropriate comments. We first show the measurement using the individual models, followed by the overall performance of the ensemble of multiple models for multi-label classification task.

\BfPara{(1) Toxic Comments} 
We measured the toxicity of YouTube comments using the toxic comments detection model. 
Figure~\ref{fig:toxic_threshold_R} shows the performance of the model in terms of TPR and TNR using different thresholds, and $0.520$ is selected as the threshold with the best trade-off. 
Applying the model on our dataset, Figure~\ref{fig:toxic_threshold_R} shows 11\% (405,290 comments) of all comments were classified as toxic.  

\BfPara{(2) Threat Comments} Similarly, the model for detecting threat comments achieved a TNR of 86\%. We set the threshold for this category to $0.220$, providing the best trade-off with a TPR of 85\% as shown in Figure~\ref{fig:threat_R}.
Adopting the model to detect threat comments, 2\% of the comments (63,939 comments) were labeled as threat.

\BfPara{(3) Insult Comments}
The insult comments model provides a TPR of 66\% and TNR of 85\%. Figure~\ref{fig:insult_R} shows the results using different thresholds. In our design, we selected $0.210$ as a threshold for predicting insult comments. Using this model with the adopted threshold, 7\% of the collected comments are detected as an insult (262,934 comments).

\BfPara{(4) Obscene Comments}
The obscene comments model, operating with a prediction threshold of $0.270$, achieves a TPR of 86\% and TNR of 88\%. Figure~\ref{fig:obscene_threshold_R} shows the results of adopting different thresholds, most of which provide high scores. Applying the model on the comments, 4\% were detected as obscene (159,823 comments).

\BfPara{(5) Identity Hate Comments}
The model for detecting identity hate comments shows a high performance as demonstrated in Figure~\ref{fig:identity_hate_R}. Using a prediction threshold of $0.140$, the achieved TPR and TNR are 74\% and 98\%, respectively.
Applying the model to YouTube comments, we found that among the comments, 3\% were labeled as identity hate comments, or 101,311 comments.

\begin{figure*}[t]
\begin{minipage}[t]{0.49\textwidth}
\centering
\includegraphics[width=0.92\textwidth]{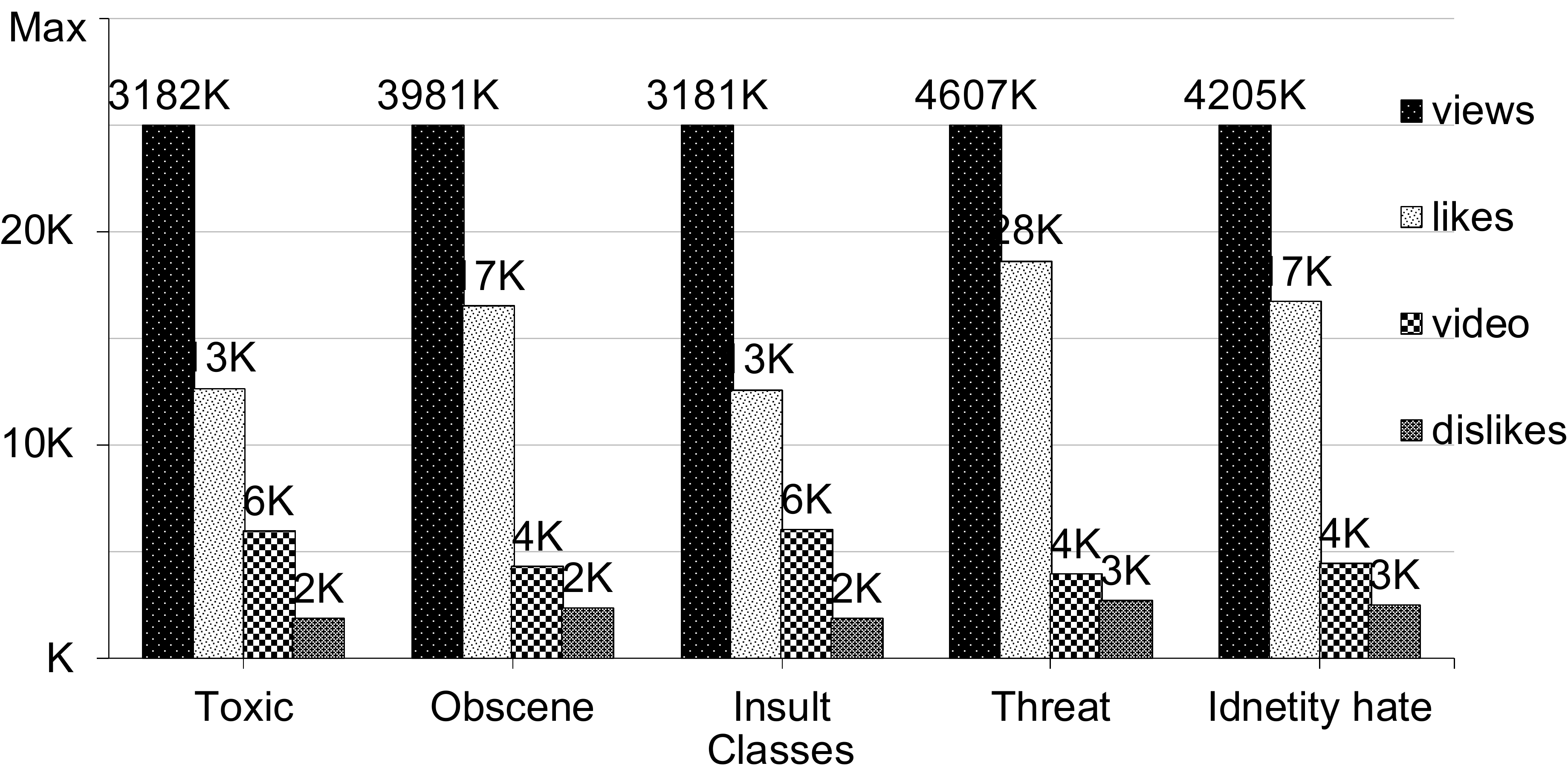}\vspace{-2mm} 
\caption{
Average number of views and likes on kids' videos containing inappropriate comments. YouTube kids' videos with unsafe comments have high number of views, likes, and dislikes.}\label{fig:classes_videos_stats}\vspace{-2mm} 
\end{minipage}%
\hfill
\begin{minipage}[t]{0.49\textwidth}
\centering
\includegraphics[width=0.92\textwidth]{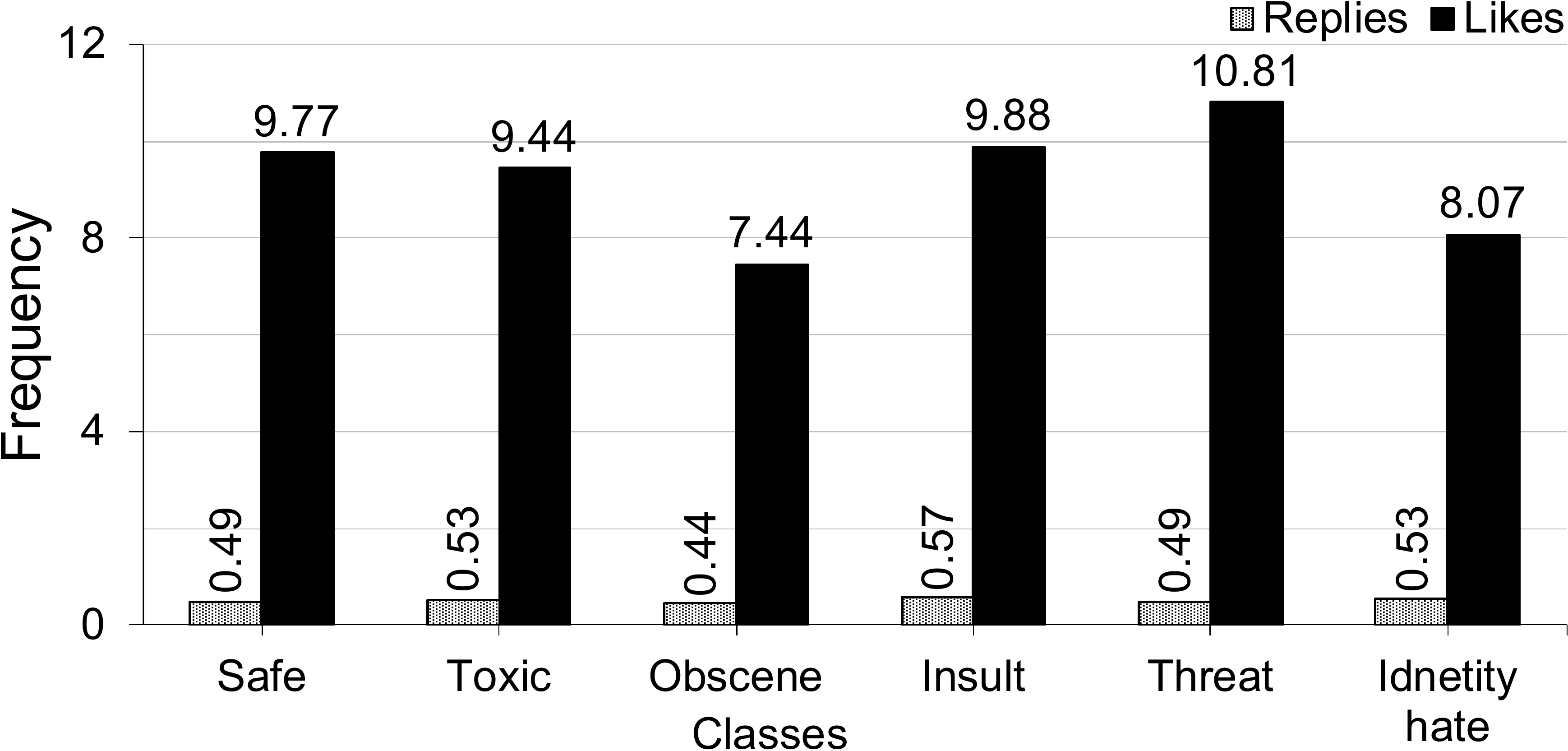}\vspace{-2mm} 
\caption{
The average number of like and replies on inappropriate comments. On average, comments associated with threat, insult and identity hate have higher number of likes and replies.}\label{fig:classes_comm_stats}\vspace{-2mm} 
\end{minipage}

\end{figure*}
\begin{figure}[t]
\includegraphics[width=0.40\textwidth]{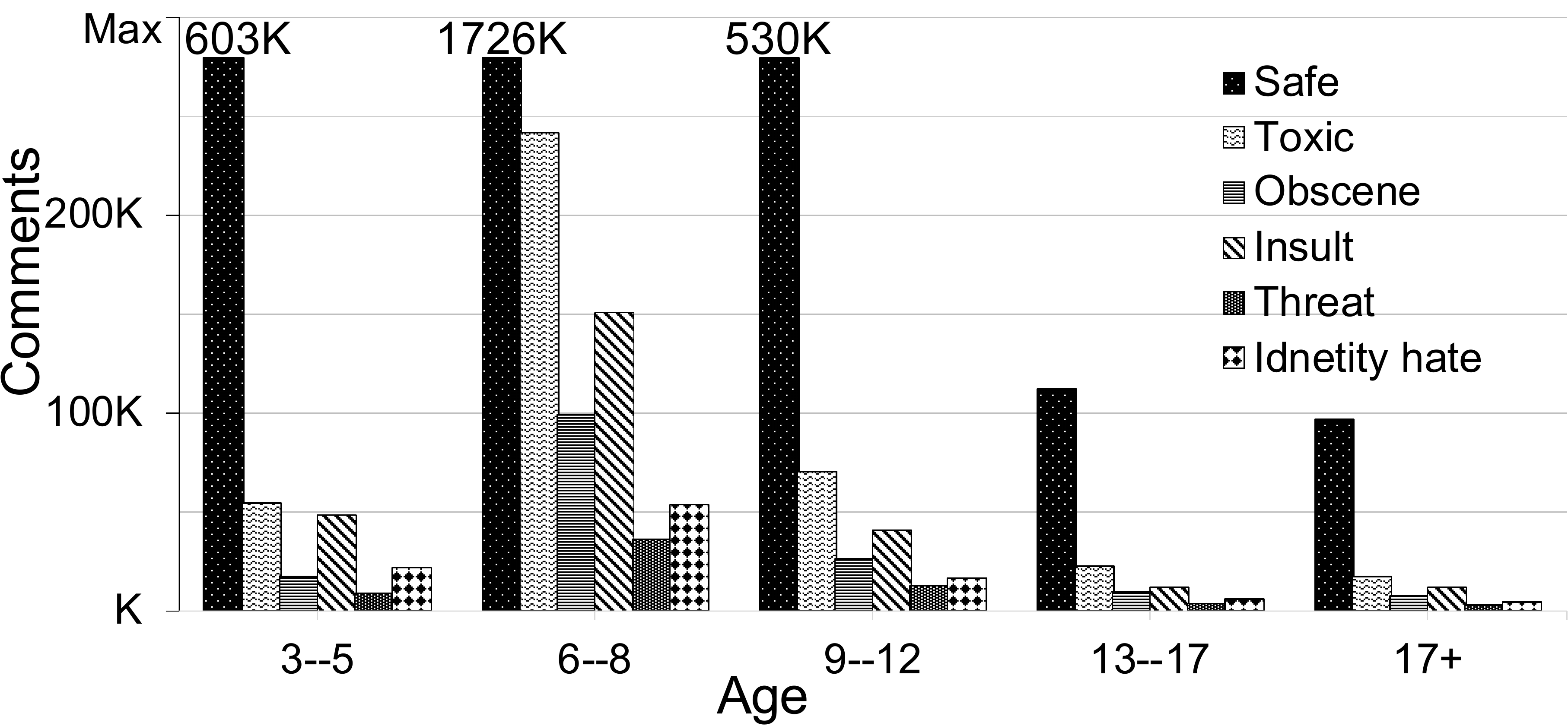}\vspace{-2mm} 
\caption{The distribution of inappropriate comments over different age groups.}\label{fig:age_classes}\vspace{-4mm} 
\end{figure}

\subsection{Inappropriateness Exposure Analysis}\label{sec:Appropriation Analysis}

\begin{table*}[t]
\caption{Distribution of the inappropriate comments over different YouTube Channels as well as the average of the inappropriate comments to the overall comments posted on each category.}\label{tab:channels_classes}\vspace{-2mm} 
\centering{
\begin{tabular}{l|c|c|c|c|c|c|c|c|c|c}
\Xhline{2\arrayrulewidth}
Channel Name            & \# Video & \# Comments & Safe   & Toxic & Obscene & Insult & Threat & \begin{tabular}[c]{@{}c@{}}Identity \\ hate\end{tabular}  & \begin{tabular}[c]{@{}c@{}}Unsafe /\\ video\end{tabular}  & \begin{tabular}[c]{@{}c@{}}Unsafe /\\ comment\end{tabular}  \\
\Xhline{2\arrayrulewidth}
Warner Bros. Pictures   & 3        & 140594      & 113569 & 20994 & 10980   & 8002   & 1761   & 2954          & 9008           & 19.2             \\
Cartoon Hangover        & 52       & 118352      & 101385 & 11013 & 4191    & 6625   & 1926   & 1664          & 326            & 14.3             \\
Talking Tom and Friends & 44       & 99293       & 88935  & 5374  & 460     & 4491   & 1077   & 2272          & 235            & 10.4             \\
Cartoon Network         & 81       & 89620       & 80142  & 4003  & 137     & 3320   & 1956   & 1763          & 117            & 10.6             \\
moviemaniacsDE          & 3        & 21052       & 11948  & 7012  & 3795    & 3631   & 595    & 1397          & 3035           & 43.2             \\
Flashback FM            & 16       & 40833       & 29301  & 8788  & 4202    & 4376   & 940    & 1170          & 721            & 28.2             \\
Mickey Mouse            & 46       & 56423       & 50159  & 2230  & 106     & 3411   & 561    & 1183          & 136            & 11.1             \\
Nickelodeon             & 38       & 46097       & 42730  & 1222  & 66      & 1629   & 489    & 595           & 89             & 7.3              \\
DEATH BATTLE!           & 3        & 45652       & 39448  & 3608  & 810     & 2242   & 1094   & 634           & 2068           & 13.6             \\
Official Pink Panther   & 60       & 41730       & 36950  & 1388  & 175     & 1738   & 353    & 2197          & 80             & 11.5          \\
\Xhline{2\arrayrulewidth}
\end{tabular}}\vspace{-2mm} 
\end{table*}

\begin{figure*}[tbh]
    \centering
    \subfigure[Safe\label{fig:clouds_safe}]
    {\includegraphics[width=0.23\textwidth]{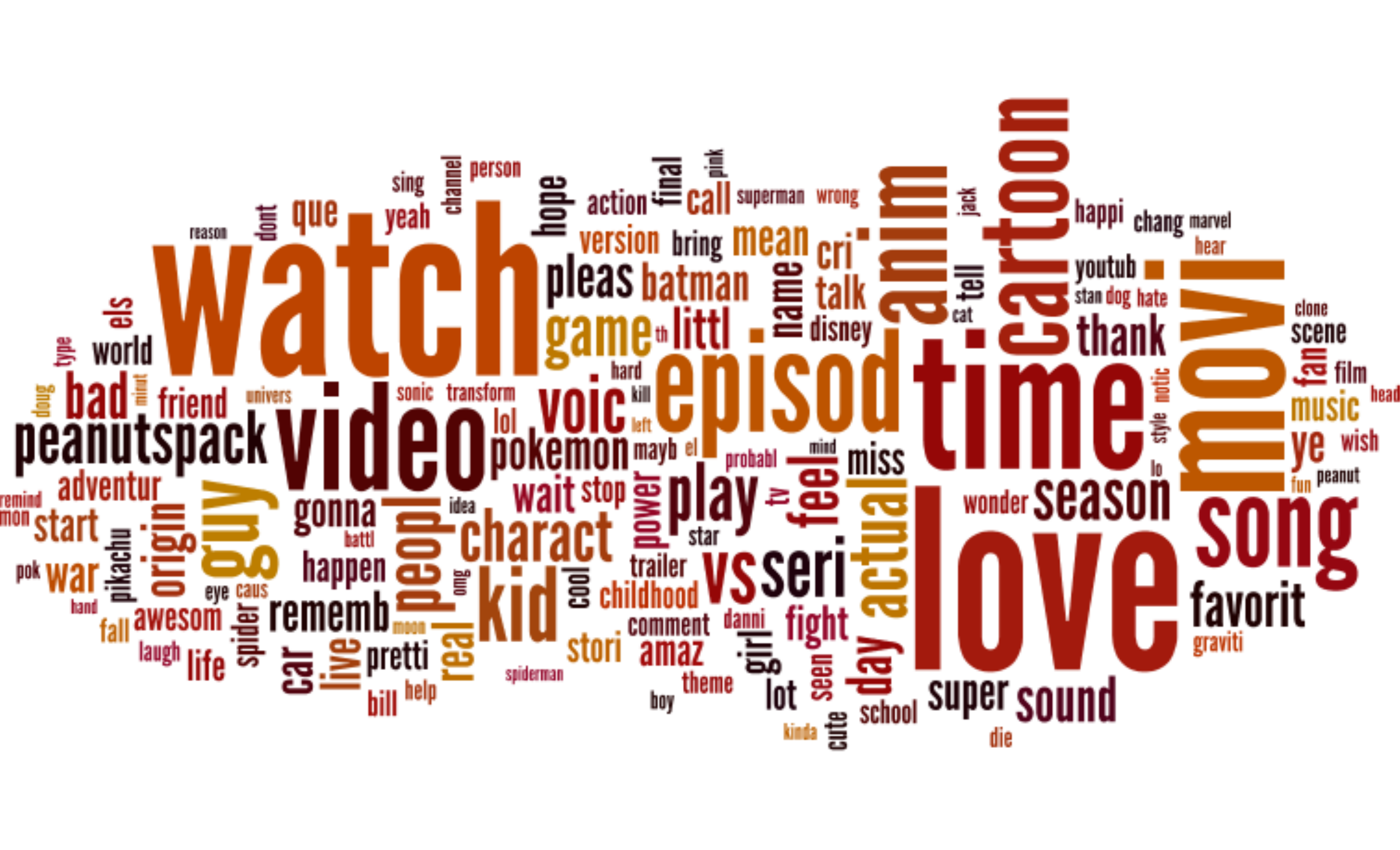}}
    \subfigure[Toxic (obscene, insult)\label{fig:clouds_toxic}] {\includegraphics[width=0.23\textwidth]{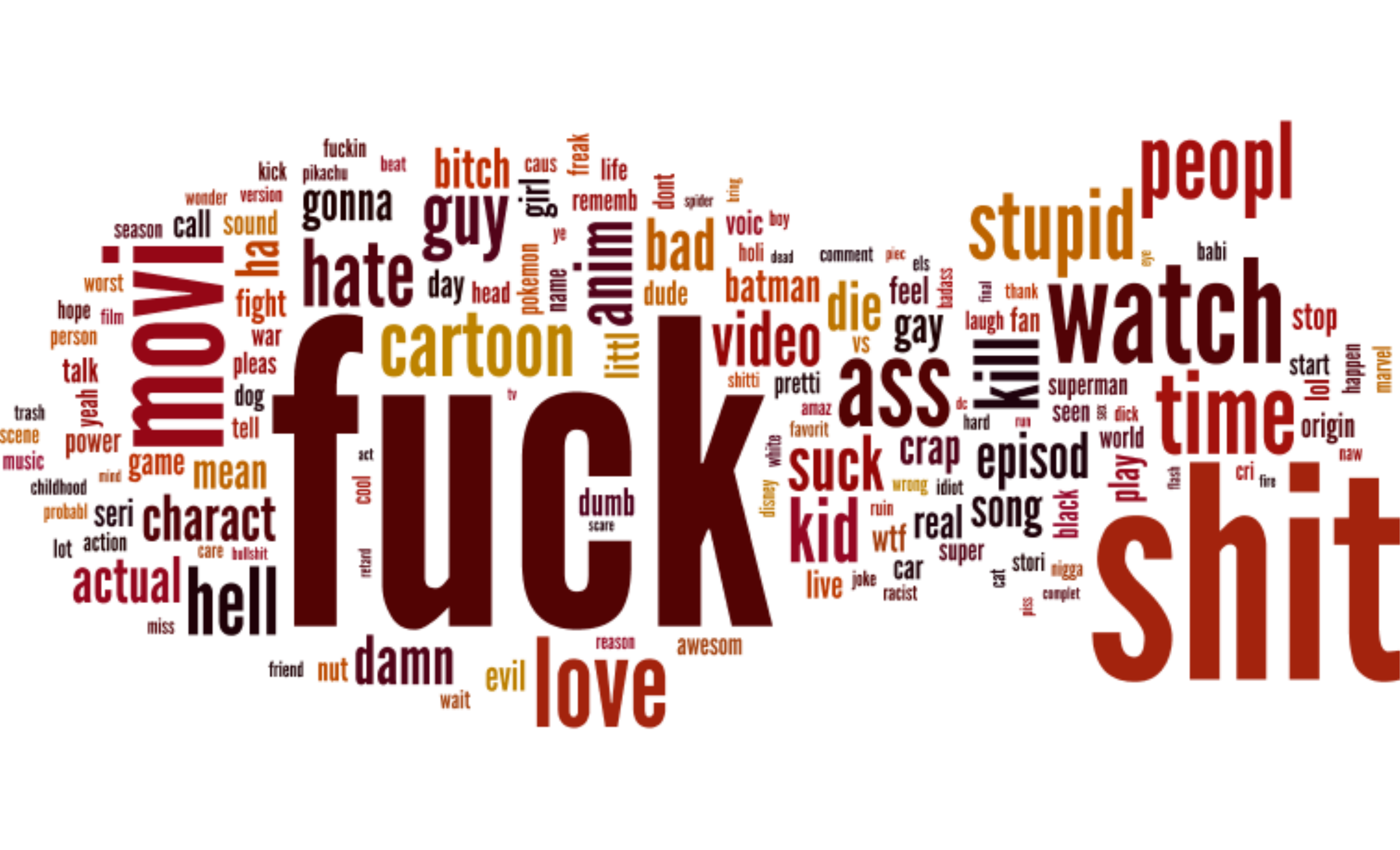}}
    \subfigure[Threat\label{fig:clouds_threat}] {\includegraphics[width=0.23\textwidth]{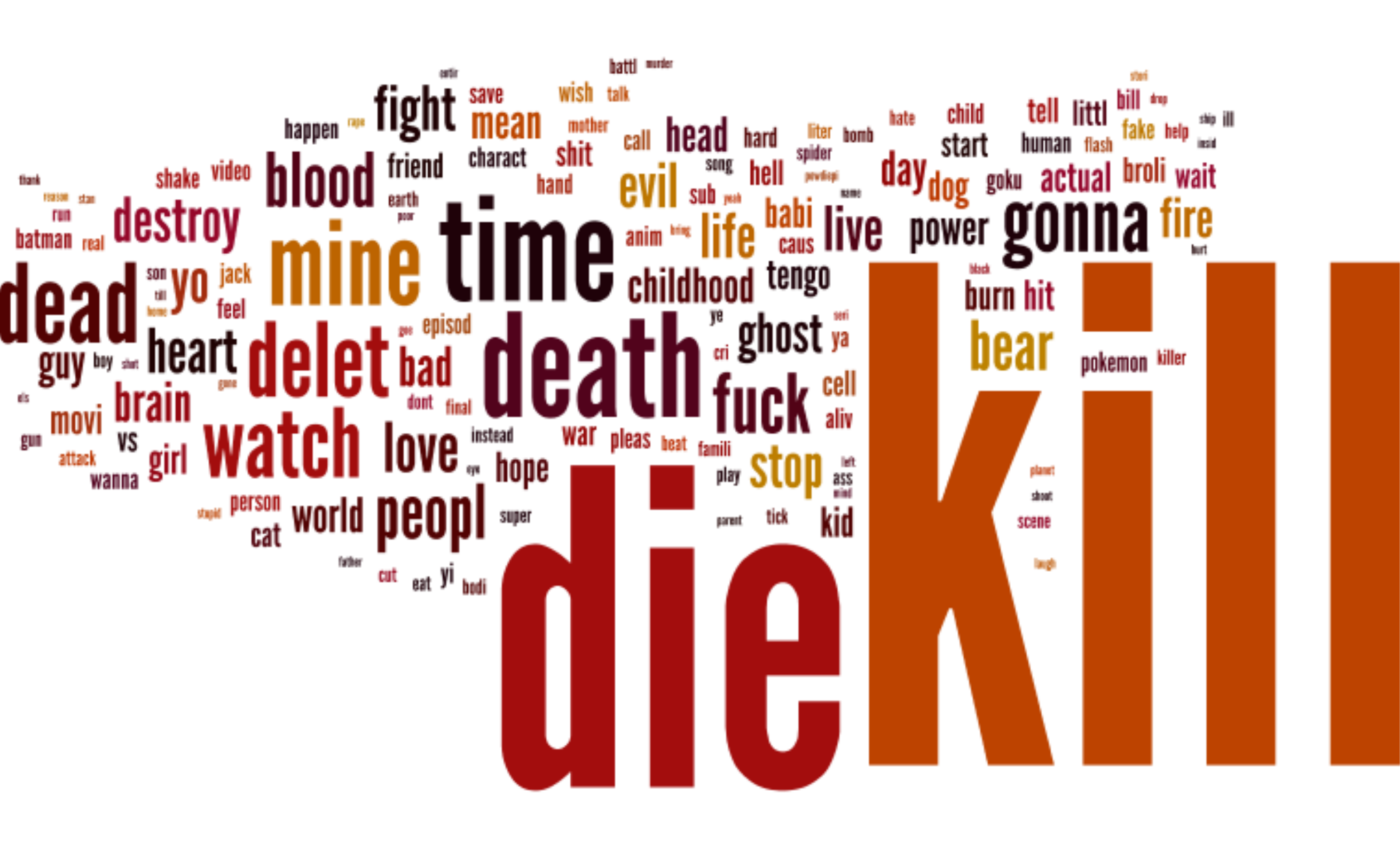}}
    \subfigure[Identity hate\label{fig:clouds_religion}] {\includegraphics[width=0.23\textwidth]{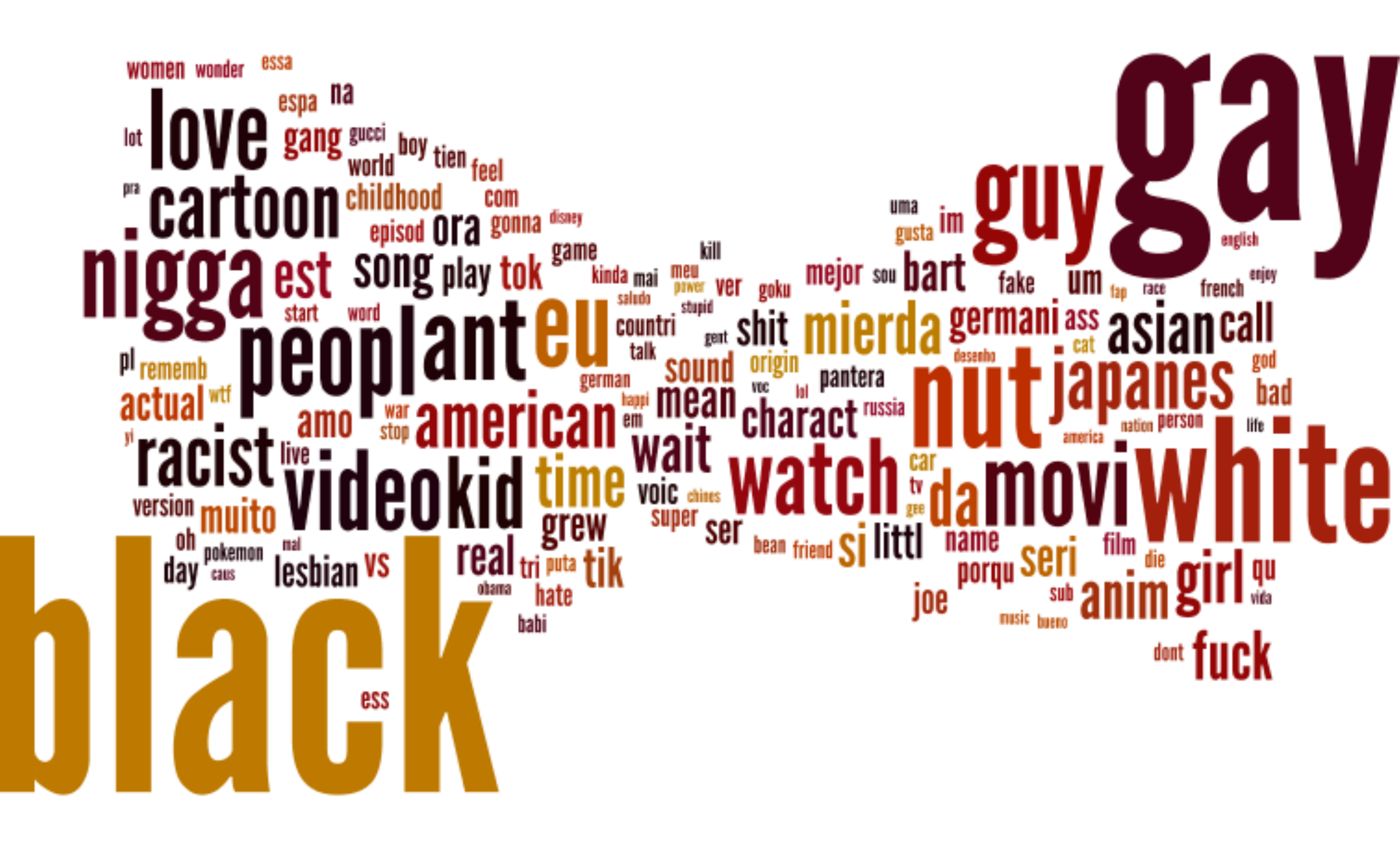}}
    \vspace{-3mm} 
    \caption{The most frequent words in YouTube comments per category. Since {\em toxic, obscene, and insult} share similar frequent words, we represented them in one cloud.}
    \label{fig:samples}\vspace{-3mm} 
\end{figure*}

\BfPara{Exposure by Age-Group}
Applying the ensemble models shows the exposure magnitude of kids to inappropriate content on YouTube comments. Investigating the exposure by different age groups,~\autoref{fig:age_classes} shows the distributions of the inappropriate comments from each age-inappropriate category over different age groups. 
For simplicity, we studied the contents of comments posted on YouTube videos targeting different age groups instead of the age in years (distributions of collected comments on videos for a specific age is shown in~\autoref{fig:age_comments}). 
Applying the ensemble models on the collected comments, we observed that toxic comments are highly common in children's videos and exceed 200,000 comments on videos only targeting the age group of six to eight years old. Insulting comments can be clearly noticed in videos targeting young children; e.g., age group 3-5 has 48,306 comments, which corresponds to 6.83\% out of the total comments collected on videos of this age group (707,161 comments). 
Comments with some sort of toxicity are also present in the collected dataset with 81,303 toxic, 17,384 obscene, 48,306 insult, 9,065 threat, and 21,329 identity hate which were detected in comments posted on videos for the age group of 3-5. These records increase to 241,352 and 36,150 for toxic and insult, respectively, on videos for the age group of 6-8. These patterns of appearance for toxic comments are observed for videos targeting all age groups.
The number of comments that contain obscene, threat, and identity hate are noticeably high for all age groups (e.g., they reach 99,165, 36,150 and 53,517, respectively, for the age group 6-8). 
We note that the reported numbers of detected categories of inappropriate comments in Figure~\ref{fig:age_classes} do not reflect their percentage with respect to the total number of comments for a certain age group. We observed that children in the age group of 3-5, which are the youngest audience, are the second most exposed to inappropriate comments, with 7.71\%, 2.46\%, 6.83\%, 1.28\%, 3.02\% for toxic, obscene, insult, threat and identity hate categories (out of the total), respectively. This age group is only second to the 13-17 age group, which has 15.54\%, 6.84\%, 7.96\%, 2.24\%, 4.20\%, for the same types.

\BfPara{Exposure and User Interaction}
Acquiring YouTube kids videos, where comments were collected and investigated, is done using the top-50 search results from the YouTube Search APIs with measures of relevance and popularity (i.e., it is safe to state that the considered videos are popular). We show statistics of users interactions with videos that contain different inappropriate content (for the five investigated categories) in~\autoref{fig:classes_videos_stats}. 
Considering the number of videos with age-inappropriate comments, we observed that the highest number of videos (6,037 videos) are reported for those with insulting comments, which has the second most number of comments among other categories (262,934). The videos with threatening comments have an average of 6.4 million views and 18,640 likes per video. More interestingly, videos with threatening comments tend to get higher interaction in terms of the number of likes (18,640) than videos with either obscene or identity hate comments (an average of 17,000 comments). 
Another observation is that the number of dislikes for the videos is positively proportional to the number of threat and identity hate comments.

We explored user interaction with inappropriate comments in terms of the number of likes and replies.
Figure \ref{fig:classes_comm_stats} shows the average number of likes and replies for comments that belong to the five inappropriate categories. The more likes and replies a comment get will increase the likelihood of that comment being shown in the top comments. We have noticed that threatening and insulting comments have the highest average of likes and replies, e.g., around 11 likes and 0.5 replies per comment for the threat category. As opposed to the other age-inappropriate categories, identity hate, and insulting comments have a high number of average replies, with an average reply of 0.53 per comment. Even though the users interaction with comments from other categories is less than threatening and insulting comments, the interaction can be seen for all categories in~\autoref{fig:classes_comm_stats}.

\BfPara{Exposure by YouTube Channel}
Investigating the top-10 most comment-contributing YouTube channels to our collected comments, \autoref{tab:channels_classes} shows the distribution of age-inappropriate comments across different channels with respect to the five investigated categories. The table highlights the number of videos of which we collected the comments as well as the number of collected comments enabling the estimation of the percentages of inappropriate comments. 
The highest number of detected inappropriate comments is reported for {\em moviemaniacsDE} channel with 43.2\% of the total comments classified as inappropriate. Furthermore, there are an alarming number of unsafe comments posted on the \textit{Warner Bros. Pictures} channel videos, where the average number of inappropriate comments is 9,008 comments per video. In contrast, \textit{Official Pink Panther} has the lowest average number of unsafe comments per video, with only 80 comments per video. 
We also observed a high number of detected inappropriate comments from the Nickelodeon channel, with 7.3\% of the total comments in this channel classified as inappropriate. This percentage was the lowest among other channels, although still an alarming score of exposure to inappropriate comments for impressionable children.

\subsection{Discussion}

\BfPara{YouTube Platform for Children}
Social media has become well-established and a part of most people's daily routine~\cite{GasserCML12}. 
Many studies have shown that children under the age of 18 spend a substantial amount of time on social media, especially on YouTube. 
A survey, conducted by the Pew Research Center in 2018, shows that 81\% of parents in the United States with children younger than 11 years of age allow their children to watch YouTube videos, and 34\% of parents stated that their children watch YouTube videos regularly~\cite{PewInternet2}.
The collection of our dataset confirms the rapid growth of popularity for YouTube videos targeting children.
More importantly, the results show that posted comments on children's videos contain contents that are inappropriate, this might affect their safety, privacy, intellect, emotion, or/and behavior.
We note that YouTube established the {\em YouTube Kids} mobile app (in February 2015) and website (in August 2019), a safe platform for kids where the comment feature is disabled. However, a large percentage of children; i.e., 80\% according to a study by~\cite{CommonSenseMedia}, still use YouTube's original website and/or mobile app. Therefore, and based on our study, children who use YouTube unsupervised might encounter inappropriate content in the comments section, highlighting the risks of media platforms, and calling for measures to ensure their safety online.

\BfPara{Awareness of Inappropriate Comments} 
This study sheds light on the exposure of adolescents to inappropriate comments on YouTube, and shows that visual and audio are not the only media that should be supervised but also the written contents. \autoref{fig:samples} shows some of the frequently inappropriate words detected to be one of the five age-inappropriate categories investigated in our study from comments posted on children's videos. 
This study shows that among inappropriate comments, there exists a large number of comments that have toxic, threatening, insulting, or/and identity hate contents which possibly can influence the psychological well-being of children.

\section{Conclusion}\label{sec:conclusion}

In this work, we studied the exposure of kids to inappropriate and comments posted on kids' YouTube videos. We studied the exposure to a five age-inappropriate categories, namely, toxic, obscene, insult, threat, and identity hate. Using an ensemble of specialized models trained on labeled data, we measured the exposure of each category by different age groups to find out that the age group of 13-17 is the most exposed group to the inappropriate comments followed by the 6-8 age group. The results show that toxic comments are common on children's videos with 10.95\% of the total comments having toxic language, followed by insults (7\%), obscene (4.32\%), identity hate (2.74\%) and threat (1.73\%) comments. We also measured users' interactions (views, likes, and dislikes) with videos having age-inappropriate comments as well as the comments themselves. We found that videos and comments with toxic or threatening comments tend to have higher interaction. 
Videos with threat comments have a high degree of popularity with an average of 4.6 million views and 28,000 likes per video. Similar popularity is observed for comments promoting identity hate with an average of 4.2 million views and 17,000 likes per video. This research shows that children are exposed to inappropriate comments, and call for increased awareness of such exposure and take measures to ensure children's safety from this exposure while on YouTube.

\BfPara{Acknowledgement} This work was supported in part by NRF grant NRF-2016K1A1A2912757 (GRL program), GPU award by NVIDIA.

\bibliographystyle{ACM-Reference-Format}
\bibliography{ref}

\end{document}